# Tracking Intrinsic Non-Hermitian Skin Effect in Lossy Lattices

Liwei Xiong,[†] Qicheng Zhang,[†] Xiling Feng, Yufei Leng, Min Pi, Shuaishuai Tong, and Chunyin Qiu[*]

Key Laboratory of Artificial Micro- and Nano-Structures of Ministry of Education
and School of Physics and Technology, Wuhan University, Wuhan 430072, China
[†]These authors contributed equally: Liwei Xiong, Qicheng Zhang
[*] To whom correspondence should be addressed: cyqiu@whu.edu.cn

*Abstract*. Non-Hermitian skin effect (NHSE), characterized by a majority of eigenstates localized at open boundaries, is one of the most iconic phenomena in non-Hermitian lattices. Despite notable experimental studies implemented, most of them witness only certain signs of the NHSE rather than the *intrinsic* exponential localization inherent in eigenstates, owing to the ubiquitous and inevitable background loss. Even worse, the experimental observation of the NHSE would be *completely obscured* in highly lossy cases. Here, we theoretically propose a dual test approach to eliminate the destructive loss effect and track the intrinsic NHSE that is essentially irrelevant to background loss. Experimentally, the effectiveness of this approach is precisely validated by one- and two-dimensional non-Hermitian acoustic lattices. Our study sheds new light on the previously untapped intrinsic aspect of the NHSE, which is of particular significance in non-Hermitian topological physics.

In last decades, non-Hermitian topological physics has been developed vigorously due to its unique properties beyond the Hermitian regime [1-5]. In particular, the non-Hermitian skin effect (NHSE), manifested as the exponential localization of extensive eigenstates at open boundaries, is a landmark phenomenon in non-Hermitian lattices [6-14]. The NHSE not only brings about a breakdown of the conventional bulk-boundary correspondence proposed in Hermitian systems [6-11], but also reflects a nontrivial point-gap topology (or associated spectral windings) unique to non-Hermitian systems [12-14]. Ever since its discovery, tremendous theoretical progress has been made in different varieties of NHSE like bipolar NHSE [15], critical NHSE [16], higher-order NHSE [17-20], geometry-dependent NHSE [21-23], etc.

Experimentally, the NHSE and associated applications have been extensively explored in various classical and quantum platforms, including electric-circuit [24-28], optical [29,30], acoustic [31-36], mechanical [37-40], quantum walks [41,42], and cold atom systems [43]. In practice, however, observing NHSE faces an inevitable constraint from the background loss (e.g., the thermal loss of material and the propagation dissipation of light or sound) [44,45]. For instance, in passive systems the NHSE is usually submerged by the loss effect (LE), despite that one may extract some NSHE information by deducting an empirical loss or using transient excitation [33,34]. By contrast, it is more accessible to observe the NHSE in active systems, since the external gains introduced by active components can counteract the LE to some extent [29,32,37-39]. Nevertheless, most of the experiments witness only some signs of the NHSE, rather than the *intrinsic* exponential localization inherent in eigenstates (which are essentially irrelevant to the *uniform* background loss). Note that it is the intrinsic NHSE that directly connects to generalized Brillouin zone and non-Bloch band topology of the non-Hermitian system. Quite specially, in electric-circuit systems one can access the intrinsic NHSE through computing the eigenmodes of an admittance matrix constructed by $N^2$ elementary impedance measurements [24-28], with $N$ characterizing the size of the system. Although this approach might be extended to other systems, it suffers from an extremely heavy workload (especially in high-dimensional systems) and is very sensitive to noise as discussed later. To sum up, it is highly desirable to develop an efficient experimental strategy to capture the intrinsic NHSE.

Here, we propose a simple but effective approach for tracking the intrinsic NHSE in lossy lattices. The key idea is intuitive: by testing a dual lattice of identical LE but reversed NHSE, the NHSE of the original lattice can be recovered by canceling the undesirable LE. This approach, dubbed dual test approach later, is demonstrated theoretically first by the paradigmatic one-dimensional (1D) Hatano-Nelson (HN) lattice model (see Fig. 1). It is then validated by our acoustic experiments, where the non-Hermitian acoustic lattice is skillfully constructed by coordinating *passive* air cavity-tube structures and *active* unidirectional couplers. This is followed by acoustic demonstrations of various two-dimensional (2D) HN lattices, which realize not only the conventional first-order edge NHSE, but also the less involved higher-order corner NHSE. Excitingly, no matter whether the NHSE or LE dominates the system, our experimental data capture precisely the intrinsic NHSE that features (loss-irrelevant) exponential sound localization. Our findings could motivate further experimental studies of non-Hermitian lattice physics, and pave the way for applications such as unidirectional sensing and energy harvesting [24,29].

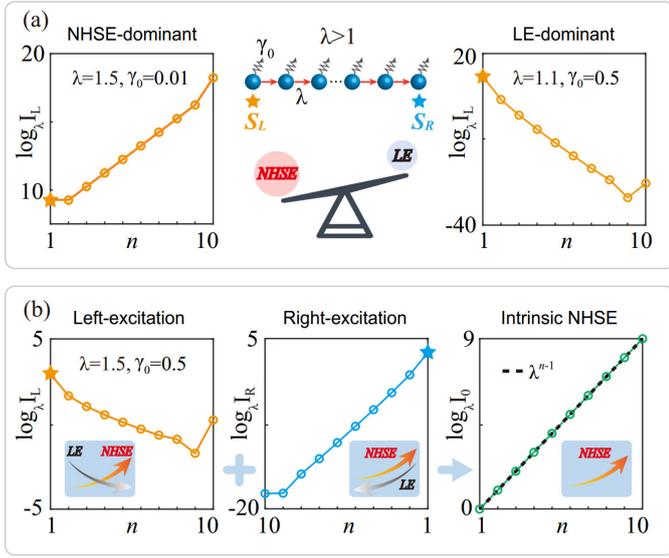

FIG. 1. Tracking intrinsic NHSE in lossy lattices. (a) Competition between the NHSE and LE, illustrated by a 1D HN model (middle) with additional uniform onsite loss $\gamma_0$. For clarity, the red arrows visualize a rightward NHSE under the nonreciprocity $\lambda > 1$. The intensity distributions in the left and right panels exemplify the NHSE- and LE-dominant phenomena, respectively, which are excited by a source (star) locating at the left end of the lattice. (b) Dual test approach used for tracking the intrinsic NHSE. The left and middle panels show the intensity distributions excited at the left and right ends, respectively. In both the cases, the lattice sites are numbered from the source. Right: the intrinsic NHSE recovered according to Eq. (4), which features exactly the exponential accumulation $I_0(n) = \lambda^{n-1}$ (dashed line). The insets sketch an intuitive understanding of our approach: the left- and right-excited states experience an identical LE but a reversed NHSE away from the source, and thus the destructive LE can be eliminated by a renormalization.

For simplicity, we start with the lossless 1D HN model with non-Hermitian Hamiltonian $\mathbf{H}$. As depicted in the middle panel of Fig. 1(a), the model features a nonreciprocal coefficient $\lambda = t_R/t_L$, with $t_L$ and $t_R$ being the leftward and rightward hoppings, respectively. Without losing generality, we assume $t_L = 1$ and focus on the rightward NHSE with $\lambda > 1$. Mathematically, the NHSE can be succinctly characterized in terms of the exponential localization of probability density, i.e., $\rho_n = \sum_m |\langle n|\psi_m^R\rangle|^2 = \lambda^{n-1}$. Here $|\psi_m^R\rangle$ is the $m$-th right eigenstate of the non-Hermitian Hamiltonian $\mathbf{H}$ and $\langle n|$ is the site vector (see *Supplemental Material* for derivation). When considering an extra *uniform* onsite loss $\gamma_0$, the eigenstates and the consequent NHSE *preserve*, although the eigenvalues are shifted by $-i\gamma_0$. Remarkably different from the eigen problem, however, the LE of $\gamma_0$ will come into play and compete with the NHSE when dealing with a practical excitation problem. This is the most common scenario encountered in real experiments.

To concretize the competition between the NHSE and LE, we consider a finite sample of $N = 10$ lattice sites and place a source at its left end to ignite the desired rightward NHSE. According to the established coupled-mode theory [46], the spectral response of the lattice reads

$$\mathbf{Q}_L \mathbf{A}_L = \mathbf{P}, \quad (1)$$

where $\mathbf{Q}_L = -i\mathbf{H} + i(\omega + i\gamma_0)\mathbf{I}$ is the dynamic matrix and $\mathbf{A}_L = (a_1^L, a_2^L, \ldots, a_N^L)^T$ is the state vector excited by the source vector $\mathbf{P} = (1,0,\ldots,0)^T$, with $\omega$ being a real-valued excitation frequency and $\mathbf{I}$ being an identity matrix of $N \times N$. This gives rise to the site-resolved intensity $I_L(n) = \langle |a_n^L|^2 \rangle$, where $\langle \cdot \rangle$ denotes an integral over frequency. In Fig. 1(a) we exemplify the intensity distributions for two typical systems, one for the case of strong nonreciprocity ($\lambda = 1.5$) but weak loss ($\gamma_0 = 0.01$), and the other for the case of strong loss ($\gamma_0 = 0.5$) but weak nonreciprocity ($\lambda = 1.1$). Log-plot is used for better characterizing the localization behavior. It shows that in the former case (left panel), although a clean exponential growth is not observable, the left-excited intensity $I_L$ accumulates rapidly toward the right end, as a hallmark of the NHSE. Conversely, for the latter case (right panel), $I_L$ is localized around the source, indicating a LE-dominant phenomenon. Note that in both systems, the irregular kinks near the lattice terminations can be attributed to the interaction of the NSHE and finite-size effect.

One question arises naturally: could we track the loss-irrelevant intrinsic NHSE from real pumping-probe experiments? Notice that an identical LE but a reversed NHSE will happen in a dual lattice of Hamiltonian $\mathbf{H}^T$. Intuitively, a pure NHSE could be extracted by canceling the LE embedded in the original and dual lattices. For the 1D HN model, the dual lattice can be realized by simply relocating the source to the right-end and renumbering the lattice sites from the source. The spectral response to the right-end excitation reads

$$\mathbf{Q}_R \mathbf{A}_R = \mathbf{P}, \quad (2)$$

where $\mathbf{Q}_R = -i\mathbf{H}^T + i(\omega + i\gamma_0)\mathbf{I}$ is the associated dynamic matrix and $\mathbf{A}_R = (a_1^R, a_2^R, \ldots, a_N^R)^T$ is the state vector excited by the source vector $\mathbf{P} = (1,0,\ldots,0)^T$. Notice that the dynamic matrices $\mathbf{Q}_L$ and $\mathbf{Q}_R$ have similarity transformations $\mathbf{Q}_L = \mathbf{\Lambda}\mathbf{M}\mathbf{\Lambda}^{-1}$ and $\mathbf{Q}_R = \mathbf{\Lambda}^{-1}\mathbf{M}\mathbf{\Lambda}$ (see *Supplemental Material*). This gives $\mathbf{M}\mathbf{\Lambda}^{-1}\mathbf{A}_L = \mathbf{\Lambda}^{-1}\mathbf{P}$ and $\mathbf{M}\mathbf{\Lambda}\mathbf{A}_R = \mathbf{\Lambda}\mathbf{P}$ according to Eqs. (1) and (2). Together with the detailed form of $\mathbf{\Lambda} = \text{diag}(1, \sqrt{\lambda}, \ldots, \sqrt{\lambda^{N-1}})$, one may derive the relation between the left- and right-excited states

$$a_n^L/a_n^R = \lambda^{n-1}. \quad (3)$$

It unveils directly the intrinsic NHSE $\rho_n = \lambda^{n-1}$, which is apparently irrelevant to the background loss $\gamma_0$. One can also track the intrinsic NHSE through the renormalized intensity distribution

$$I_0(n) = \sqrt{I_L(n)/I_R(n)} = \lambda^{n-1}. \quad (4)$$

Here the frequency-integrated quantities $I_L(n)$ and $I_R(n)$ are employed to reduce the error in real experiments.

To verify the effectiveness of our dual test approach, we consider a concrete case with $\lambda = 1.5$ and $\gamma_0 = 0.5$. As shown in Fig. 1(b), under the competition between the NHSE and LE, the left-excited intensity distribution $I_L$ (left panel) exhibits an overall attenuation away from the source, except for the kink at the tail. In contrast, the right-excited intensity distribution $I_R$ (middle panel) shows an even sharper decay due to the double kill of the LE and reversed NHSE. Excitingly, the renormalized intensity $I_0 = \sqrt{I_L/I_R}$ (right panel, circles) reproduces exactly the intrinsic NHSE depicted by $\rho_n = \lambda^{n-1}$ (dashed line), which grows exponentially toward the right end of the lattice.

It is worth pointing out that our dual test approach is applicable to any non-Hermitian lattice, provided that its Hamiltonian $\mathbf{H}$ satisfies $\mathbf{H} = \mathbf{\Lambda}\widetilde{\mathbf{H}}\mathbf{\Lambda}^{-1}$. Here, $\widetilde{\mathbf{H}}$ is a Hermitian Hamiltonian and $\mathbf{\Lambda}$ is a diagonal matrix that signifies the NHSE. Physically, as long as a dual lattice (of Hamiltonian $\mathbf{H}^T$) with reversed couplings and identical onsite loss is fabricated, a purely intrinsic NHSE can be retrieved by the renormalization protocol (see more details in *Supplemental Material*). It is of interest that in many cases, the dual lattice is exactly equivalent to its original lattice with relocated source and renumbered sites, such as the 1D and 2D HN lattices involved in our experiments below. This enables us to track the intrinsic NHSE simply from a single sample.

Below, we present experimental evidence for our dual test approach. As shown in Fig. 2(a), we consider first an acoustic emulation of the 1D HN lattice. The experimental setup comprises six identical air-filled cavities and five sets of active unidirectional couplers. The air cavities have a uniform dipole resonance frequency $\omega_0 = 5080$ Hz and an inherent loss $\gamma_0 = 40.5$ Hz. The narrow tubes between two adjacent cavities generate a fixed reciprocal coupling $t_0 = 83.2$ Hz. The unidirectional couplers [32,47] realize a controllable unidirectional coupling $\kappa$, whose value is tuned to 57.9 Hz, 37.1 Hz, and 17.0 Hz, respectively, enabling the nonreciprocity $\lambda = 1 + \kappa/t_0 = 1.70$, 1.45, and 1.20 in the following experiments. All unidirectional couplings are calibrated by a two-cavity system displayed in Fig. 2(b). More experimental details can be found in *Supplemental Material*.

Experimentally, we place first a point-like sound source at the leftmost cavity of the sample, and detect the pressure signal $p_n^L(\omega)$ at each cavity $n$. [Figure 2(c) exemplifies the site-resolved spectral response measured for the case of $\lambda = 1.20$] This further gives the frequency-integrated acoustic intensity $I_L(n) = \langle |p_n^L|^2 \rangle$. As shown in Fig. 2(e), for the system with a relatively large nonreciprocity ($\lambda = 1.70$), the acoustic intensity $I_L$ amplifies gradually toward the right, indicating a desired rightward NHSE signal. As $\lambda$ is reduced to 1.45, the LE becomes more competitive with the NHSE, leading to comparable acoustic intensities at the two ends of the acoustic lattice. For the case of $\lambda = 1.20$, the acoustic intensity behaves an overall decay away from the source, and the weakened rightward NHSE is completely covered up by the LE. To recover the NHSE, we relocate the sound source to the right end and collect the frequency-integrated acoustic intensity $I_R$ [Fig. 2(f)]. As expected, all experimental data manifest a fast decay away from the source, owing to the synergistic effect of the LE and reversed NHSE. Figure 2(g) shows the renormalized acoustic intensity $I_0 = \sqrt{I_L/I_R}$ (symbols) for the three different nonreciprocities. Remarkably, all of them exhibit a nearly perfect exponential growth toward the right end of the sample, even for the case of $\lambda = 1.20$ where the NHSE seems lost in loss. The effectiveness of our dual test approach can be seen clearly from the precise agreement with the theoretical prediction (dashed line).

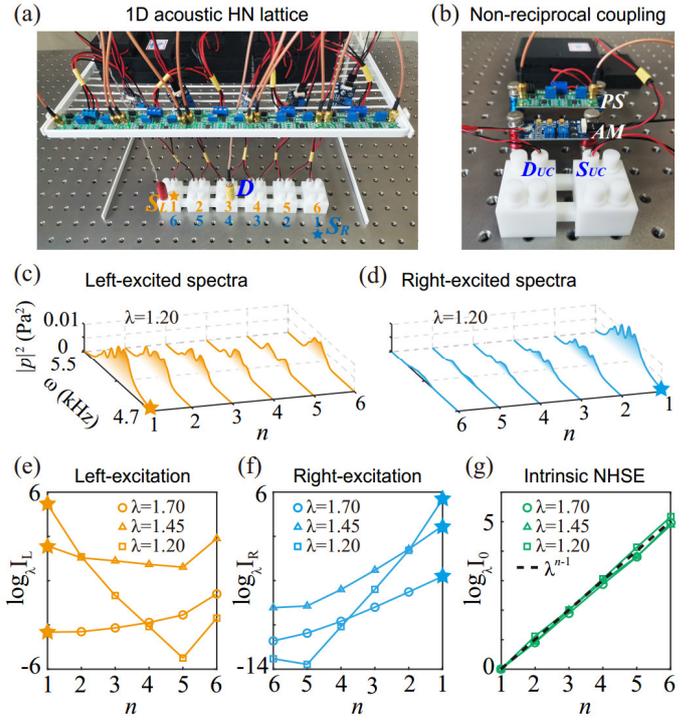

FIG. 2. Experimental evidence for the 1D HN lattice. (a) Acoustic emulation of the 1D HN lattice, which consists of a passive cavity-tube structure decorating with active unidirectional couplers. $S_L$ ($S_R$) and $D$ represent a left-excited (right-excited) source and a detector, respectively. (b) Nonreciprocal coupling between adjacent cavities, realized by two connecting tubes plus a fully controllable unidirectional coupler. The latter consists of a microphone $D_{UC}$, an amplifier $AM$, a phase shifter $PS$, and a loudspeaker $S_{UC}$. (c), (d) Site-resolved spectra measured for two different excitations. Both exhibit an overall decay away from the source and fail to display the NHSE directly. (e), (f) Frequency-integrated acoustic intensity distributions for three different nonreciprocities. (g) Renormalized acoustic intensity from $I_L$ and $I_R$. Precisely, all experimental data (symbols) reproduce well the theoretically predicted intrinsic NHSE (dashed line).

Next, we proceed to unveil the intrinsic NHSE in higher-dimensional systems. Figure 3(a) depicts a 2D version of the HN model with nonreciprocal coefficients $\lambda_x \geq 1$ and $\lambda_y \geq 1$ in two directions. The associated intrinsic NHSE can be characterized by the exponential distribution $\lambda_x^{n_x-1}\lambda_y^{n_y-1}$ (see *Supplemental Material*). The model can be emulated in acoustics with a 4×4 cavity-tube structure plus 24 unidirectional couplers, as shown in Fig. 3(b). The reciprocal couplings realized by narrow tubes are $t_x = 43.3$ Hz and $t_y = 34.1$ Hz, while the unidirectional couplings $\kappa_x$ and $\kappa_y$ are tuned to realize the nonreciprocities $\lambda_x = 1 + \kappa_x/t_x$ and $\lambda_y = 1 + \kappa_y/t_y$ in the $x$ and $y$ directions, respectively. By controlling the values of $\lambda_x$ and $\lambda_y$, we can achieve both the first-order and higher-order NHSEs [Figs. 3(c)-3(e)].

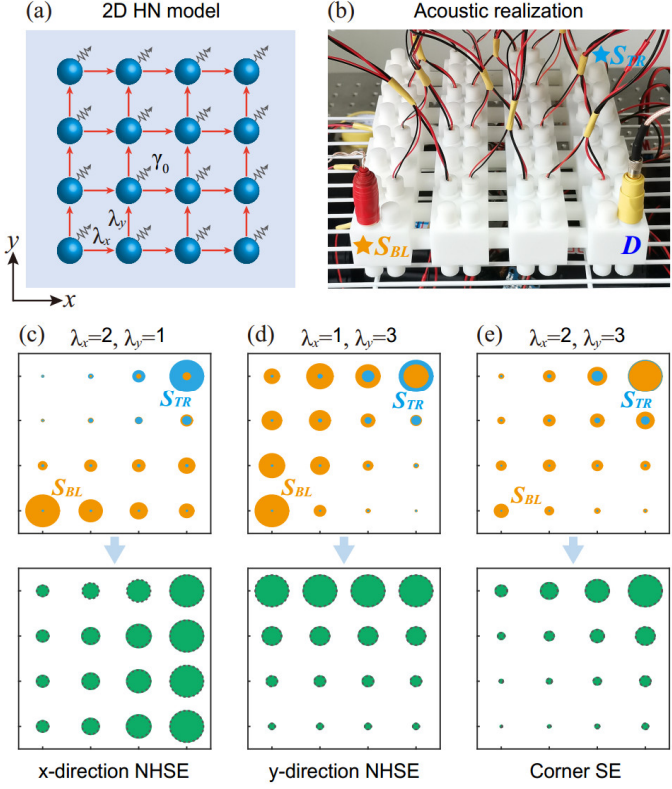

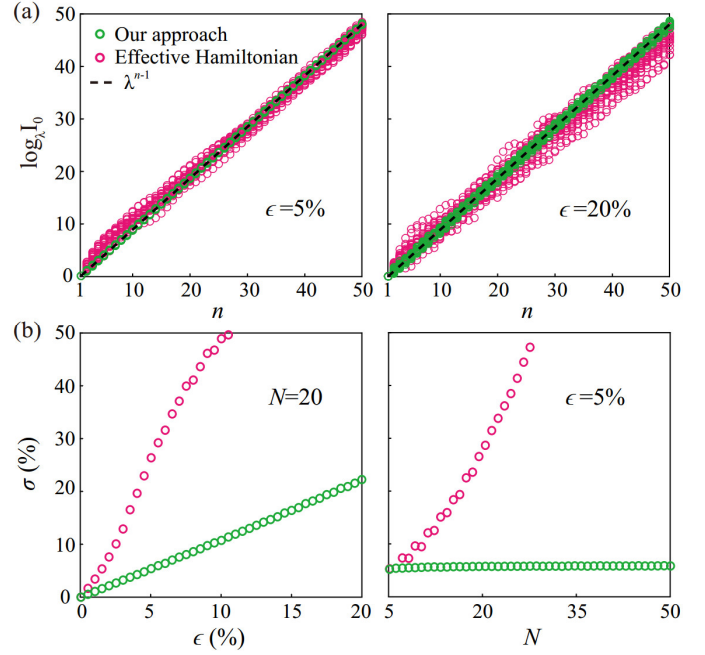

FIG. 3. Extension to 2D systems. (a) 2D HN model characterized by the nonreciprocal coefficients $\lambda_x$ and $\lambda_y$, plus a uniform onsite loss $\gamma_0$. (b) Acoustic realization of the 2D lattice, where $S_{BL}$ ($S_{TR}$) and $D$ represent a source at the bottom-left (top-right) cavity and a movable detector, respectively. (c)-(e) Experimental results for three sets of different nonreciprocities, exhibiting the $x$-direction, $y$-direction, and corner NHSEs, respectively. Top panels: acoustic intensity fields under the $S_{BL}$ and $S_{TR}$ excitations, characterized by the areas of the orange and blue disks, respectively. Bottom panels: the intrinsic NHSEs (green disks) tracked by our dual test approach, in comparison to their theoretical predictions (black dashed circles).

We consider first the case of $\lambda_x = 2$ and $\lambda_y = 1$, which exhibits nonreciprocity in the $x$ direction only. The acoustic lattice is excited successively from the bottom-left and top-right cavities, and the corresponding acoustic intensity distributions are shown by orange and blue disks in Fig. 3(c) (top panel). As expected, neither of them reflects the rightward NHSE (precited by $\lambda_x > 1$), since both the fields are localized around the source. However, as shown in the bottom panel of Fig. 3(c), the renormalized intensity distribution visualizes clearly a uniform increase toward the right edge, a manifestation of the first-order $x$-direction NHSE. The experimental data (green disks) exhibit a perfect agreement with the theoretical results (black dashed circles). Similarly, the intrinsic first-order $y$-direction NHSE is exemplified by setting $\lambda_x = 1$ and $\lambda_y = 3$ [Fig. 3(d)]. Comparing to the above $x$-direction NHSE, the $y$-direction NHSE exhibits a more pronounced localization due to the stronger nonreciprocity in the $y$ direction. More importantly, when we tune the nonreciprocities to $\lambda_x = 2$ and $\lambda_y = 3$, the exponential localization occurs in both directions, manifesting the corner NHSE in Fig. 3(e). Such a higher-order NHSE in the simplest 2D HN model [6,25] has not been experimentally demonstrated before.

FIG. 4. Robustness analysis. (a) Intrinsic NHSEs tracked by our dual test approach under the stochastic measurement perturbations $\epsilon = 5\%$ (left) and $20\%$ (right), comparing with those unveiled by effective Hamiltonian approach. Here 20 independent configurations are demonstrated. (b) Dependences of the global error $\sigma$ on the random perturbation $\epsilon$ (left) and the site number $N$ (right). The ensemble average is performed over 5000 configurations. $\lambda = 2.0$ and $\gamma_0 = 0.35$ are involved in the above simulation experiments.

As demonstrated by our above experiments, we have precisely captured the intrinsic NHSE in 1D and 2D acoustic lattices. Notice that non-Hermitian eigenstates have been measured in electric-circuit lattices [24-28], in which an admittance matrix is constructed by $N^2$ elementary impedance measurements. Similar scheme could also be applied in acoustic systems (see *Supplemental Material*). Using the retrieved eigenstates, this provides another way (dubbed effective Hamiltonian approach) to characterize the intrinsic NHSE. As compared in Fig. 4, except for the remarkably released workload, our dual test approach shows a much stronger robustness against random perturbations that are inevitable in experiments. To demonstrate this, we implement the following simulation experiments to 1D HN model, in which a random perturbation of $[-\epsilon, \epsilon]$ appears in each pressure signal measurement. Figure 4(a) shows the data retrieved for 20 independent experiments under $\epsilon = 5\%$ and $\epsilon = 20\%$. Clearly, all data retrieved by our dual test approach (green circles) concentrate to the prediction $\rho_n = \lambda^{n-1}$ (dashed line), in contrast to those dispersed data achieved by the effective Hamiltonian approach (magenta circles). To quantitatively describe the deviations, we define a global error $\sigma = \sqrt{N^{-1}\Sigma_n \rho_n^{-2}[I_e(n) - \rho_n]^2}$, where $\rho_n = \lambda^{n-1}$ characterizes the intrinsic NHSE, and $I_e$ represents the renormalized intensity in our approach or the probability density retrieved from the effective Hamiltonian approach. The left panel of Fig. 4b presents the dependence of global error $\sigma$ on the random perturbation $\epsilon$ for a size-fixed sample ($N = 20$), implemented for an ensemble average over 5000 configurations. It shows that the global error $\sigma$ of our approach behaves nearly linear with $\epsilon$, and exhibits a much slower growth than that of the effective Hamiltonian approach. More importantly, as exemplified by the fixed random error $\epsilon = 5\%$ (Fig. 4b, right panel), the global error of our approach remains nearly constant as the site number $N$ increases, in contrast to the sharp increase manifested in the effective Hamiltonian approach. This is mainly because our approach does not involve a matrix inversion that is embedded in the effective Hamiltonian approach (see Supplemental Material).

To conclude, we have theoretically proposed and experimentally verified a dual test approach for tracking the intrinsic NHSE in lossy lattices. The robustness of our approach has been clearly unveiled through a global error analysis, by contrast to the effective Hamiltonian approach. Our approach can be extended to study many other distinctive NHSE phenomena, such as dislocation NHSE [48,49] and gauge field-enriched NHSE [50-52], whose observations will also be hindered by background loss. Our approach can even be applied to those non-Hermitian models with extra lattice symmetry and topologically nontrivial edge states. This is exemplified by the paradigmatic non-Hermitian Su-Schrieffer-Heeger model [6] (see Supplemental Material). Resorting to the intrinsic NHSE, which is closely related to the generalized Brillouin zone and non-Bloch topological invariants, one can further demonstrate the critical aspects of non-Hermitian topological band theory (like non-Bloch bulk-boundary correspondence [6,9,10]).

## ACKNOWLEDGEMENTS

This work is supported by the National Natural Science Foundation of China (Grant No. 11890701, 12104346, 11674250), the Young Top-Notch Talent for Ten Thousand Talent Program (2019-2022).

## References

[1] Y. Ashida, Z. Gong, and M. Ueda, Non-Hermitian physics, Adv. Phys. **69**, 249 (2020).
[2] E. J. Bergholtz, J. C. Budich, and F. K. Kunst, Exceptional topology of non-Hermitian systems, Rev. Mod. Phys. **93**, 015005 (2021).
[3] K. Ding, C. Fang, and G. Ma, Non-Hermitian topology and exceptional-point geometries, Nat. Rev. Phys. **4**, 745 (2022).
[4] X. Zhang, T. Zhang, M.-H. Lu, and Y.-F. Chen, A Review on Non-Hermitian Skin Effect, Adv. Phys. X **7**, 2109431 (2022).
[5] N. Okuma and M. Sato, Non-Hermitian topological phenomena: A review, Annual Review of Condensed Matter Physics **14**, 83 (2023).
[6] S. Yao and Z. Wang, Edge States and Topological Invariants of Non-Hermitian Systems, Phys. Rev. Lett. **121**, 086803 (2018).
[7] V. M. Martinez Alvarez, J. E. Barrios Vargas, and L. E. F. Foa Torres, Non-Hermitian robust edge states in one dimension: Anomalous localization and eigenspace condensation at exceptional points, Phys. Rev. B **97**, 121401(R) (2018).
[8] H. Shen, B. Zhen, and L. Fu, Topological Band Theory for Non-Hermitian Hamiltonians, Phys. Rev. Lett. **120**, 146402 (2018).
[9] F. K. Kunst, E. Edvardsson, J. C. Budich, and E. J. Bergholtz, Biorthogonal Bulk-Boundary Correspondence in Non-Hermitian Systems, Phys. Rev. Lett. **121**, 026808 (2018).
[10] K. Yokomizo and S. Murakami, Non-Bloch Band Theory of Non-Hermitian Systems, Phys. Rev. Lett. **123**, 066404 (2019).
[11] C. H. Lee and R. Thomale, Anatomy of skin modes and topology in non-Hermitian systems, Phys. Rev. B **99**, 201103 (2019).
[12] N. Okuma, K. Kawabata, K. Shiozaki, and M. Sato, Topological Origin of Non-Hermitian Skin Effects, Phys. Rev. Lett. **124**, 086801 (2020).
[13] K. Zhang, Z. Yang, and C. Fang, Correspondence between Winding Numbers and Skin Modes in Non-Hermitian Systems, Phys. Rev. Lett. **125**, 126402 (2020).
[14] D. S. Borgnia, A. J. Kruchkov, and R.-J. Slager, Non-Hermitian Boundary Modes and Topology, Phys. Rev. Lett. **124**, 056802 (2020).
[15] F. Song, S. Yao, and Z. Wang, Non-Hermitian Topological Invariants in Real Space, Phys. Rev. Lett. **123**, 246801 (2019).
[16] L. Li, C. H. Lee, S. Mu, and J. Gong, Critical non-Hermitian skin effect, Nat. Commun. **11**, 5491 (2020).
[17] C. H. Lee, L. Li, and J. Gong, Hybrid Higher-Order Skin-Topological Modes in Nonreciprocal Systems, Phys. Rev. Lett. **123**, 016805 (2019).
[18] K. Kawabata, M. Sato, and K. Shiozaki, Higher-order non-Hermitian skin effect, Phys. Rev. B **102**, 205118 (2020).
[19] R. Okugawa, R. Takahashi, and K. Yokomizo, Second-order topological non-Hermitian skin effects, Phys. Rev. B **102**, 241202 (2020).
[20] Y. Li, C. Liang, C. Wang, C. Lu, and Y.-C. Liu, Gain-loss-induced hybrid skin-topological effect, Phys. Rev. Lett. **128**, 223903 (2022).
[21] X.-Q. Sun, P. Zhu, and T. L. Hughes, Geometric Response and Disclination-Induced Skin Effects in Non-Hermitian Systems, Phys. Rev. Lett. **127**, 066401 (2021).
[22] K. Zhang, Z. Yang, and C. Fang, Universal non-Hermitian skin effect in two and higher dimensions, Nat. Commun. **13**, 1 (2022).
[23] K. Zhang, C. Fang, and Z. Yang, Dynamical Degeneracy Splitting and Directional Invisibility in Non-Hermitian Systems, arXiv:2211.07783 (2023).
[24] T. Helbig, T. Hofmann, S. Imhof, M. Abdelghany, T. Kiessling, L. W. Molenkamp, C. H. Lee, A. Szameit, M. Greiter, and R. Thomale, Generalized bulk–boundary correspondence in non-Hermitian topolectrical circuits, Nat. Phys. **16**, 747 (2020).
[25] T. Hofmann, T. Helbig, F. Schindler, N. Salgo, M. Brzezińska, M. Greiter, T. Kiessling, D. Wolf, A. Vollhardt, A. Kabaši, C. H. Lee, A. Bilušić, R. Thomale, and T. Neupert, Reciprocal skin effect and its realization in a topolectrical circuit, Phys. Rev. Research **2**, 023265 (2020).
[26] S. Liu, R. Shao, S. Ma, L. Zhang, O. You, H. Wu, Y. J. Xiang, T. J. Cui, and S. Zhang, Non-Hermitian skin effect in a non-Hermitian electrical circuit, Research **2021**, 5608038 (2021).
[27] D. Zou, T. Chen, W. He, J. Bao, C. H. Lee, H. Sun, and X. Zhang, Observation of hybrid higher-order skin-topological effect in non-Hermitian topolectrical circuits, Nat. Commun. **12**, 7201 (2021).
[28] P. Zhu, X.-Q. Sun, T. L. Hughes, and G. Bahl, Higher rank chirality and non-Hermitian skin effect in a topolectrical circuit, Nat. Commun. **14**, 720 (2023).
[29] S. Weidemann, M. Kremer, T. Helbig, T. Hofmann, A. Stegmaier, M. Greiter, R. Thomale, and A. Szameit, Topological funneling of light, Science **368**, 311 (2020).
[30] Y. Song, W. Liu, L. Zheng, Y. Zhang, B. Wang, and P. Lu, Two-dimensional non-Hermitian skin effect in a synthetic photonic lattice, Phys. Rev. Appl. **14**, 064076 (2020).
[31] X. Zhang, Y. Tian, J.-H. Jiang, M.-H. Lu, and Y.-F. Chen, Observation of higher-order non-Hermitian skin effect, Nat. Commun. **12**, 5337 (2021).
[32] L. Zhang, Y. Yang, Y. Ge, Y. J. Guan, Q. Chen, Q. Yan, F. Chen, R. Xi, Y. Li, D. Jia, S. Q. Yuan, H. X. Sun, H. Chen, and B. Zhang, Acoustic non-Hermitian skin effect from twisted winding topology, Nat. Commun. **12**, 6297 (2021).
[33] H. Gao, H. Xue, Z. Gu, L. Li, W. Zhu, Z. Su, J. Zhu, B. Zhang, and Y. D. Chong, Anomalous Floquet non-Hermitian skin effect in a ring resonator lattice, Phys. Rev. B **106**, 134112 (2022).
[34] Z. Gu, H. Gao, H. Xue, J. Li, Z. Su, and J. Zhu, Transient non-Hermitian skin effect, Nat. Commun. **13**, 7668 (2022).
[35] Q. Zhou, J. Wu, Z. Pu, J. Lu, X. Huang, W. Deng, M. Ke, and Z. Liu, Observation of exceptional points and skin effect

correspondence in non-Hermitian phononic crystals, Nat. Commun. **14**, 4569 (2023).
- [36] T. Wan, K. Zhang, J. Li, Z. Yang, and Z. Yang, Observation of dynamical degeneracy splitting for the non-Hermitian skin effect, arXiv:2303.11109 (2023).
- [37] M. Brandenbourger, X. Locsin, E. Lerner, and C. Coulais, Non-reciprocal robotic metamaterials, Nat. Commun. **10**, 4608 (2019).
- [38] A. Ghatak, M. Brandenbourger, J. van Wezel, and C. Coulais, Observation of non-Hermitian topology and its bulk–edge correspondence in an active mechanical metamaterial, Proc. Natl. Acad. Sci. USA **117**, 29561 (2020).
- [39] W. Wang, X. Wang, and G. Ma, Non-Hermitian Morphing of Topological Modes, Nature **608**, 50 (2022).
- [40] W. Wang, M. Hu, X. Wang, G. Ma, and K. Ding, Experimental Realization of Geometry-Dependent Skin Effect in a Reciprocal Two-Dimensional Lattice, arXiv:2302.06314 (2023).
- [41] L. Xiao, T. Deng, K. Wang, G. Zhu, Z. Wang, W. Yi, and P. Xue, Non-Hermitian bulk–boundary correspondence in quantum dynamics, Nat. Phys. **16**, 761 (2020).
- [42] L. Xiao, T. Deng, K. Wang, Z. Wang, W. Yi, and P. Xue, Observation of Non-Bloch Parity-Time Symmetry and Exceptional Points, Phys. Rev. Lett. **126**, 230402 (2021).
- [43] Q. Liang, D. Xie, Z. Dong, H. Li, H. Li, B. Gadway, W. Yi, and B. Yan, Dynamic signatures of non-Hermitian skin effect and topology in ultracold atoms, Phys. Rev. Lett. **129**, 070401 (2022).
- [44] W.-T. Xue, M.-R. Li, Y.-M. Hu, F. Song, and Z. Wang, Simple formulas of directional amplification from non-Bloch band theory, Phys. Rev. B **103**, L241408 (2021).
- [45] H. Schomerus, Fundamental constraints on the observability of non-Hermitian effects in passive systems, arXiv:2207.09014 (2022).
- [46] W. Suh, Z. Wang, and S. Fan, Temporal coupled-mode theory and the presence of non-orthogonal modes in lossless multimode cavities, IEEE Journal of Quantum Electronics **40**, 1511 (2004).
- [47] Q. Zhang, Y. Li, H. Sun, X. Liu, L. Zhao, X. Feng, X. Fan, and C. Qiu, Observation of Acoustic Non-Hermitian Bloch Braids and Associated Topological Phase Transitions, Phys. Rev. Lett. **130**, 017201 (2023).
- [48] F. Schindler and A. Prem, Dislocation non-Hermitian skin effect, Phys. Rev. B **104**, L161106 (2021).
- [49] B. A. Bhargava, I. C. Fulga, J. van den Brink, and A. G. Moghaddam, Non-Hermitian skin effect of dislocations and its topological origin, Phys. Rev. B **104**, L241402 (2021).
- [50] M. Lu, X.-X. Zhang, and M. Franz, Magnetic Suppression of Non-Hermitian Skin Effects, Phys. Rev. Lett. **127**, 256402 (2021).
- [51] Y. Peng, J. Jie, D. Yu, and Y. Wang, Manipulating the non-Hermitian skin effect via electric fields, Phys. Rev. B **106**, L161402 (2022).
- [52] C.-A. Li, B. Trauzettel, T. Neupert, and S.-B. Zhang, Enhancement of Second-Order Non-Hermitian Skin Effect by Magnetic Fields, arXiv:2212.1469 (2022).